# Integration of LiDAR and multispectral images for exposure and earthquake vulnerability estimation. Application in Lorca, Spain.

Yolanda Torres[a] (y.torres@upm.es), José Juan Arranz[a] (josejuan.arranz@upm.es), Jorge M. Gaspar-Escribano[a] (jorge.gaspar@upm.es), Azadeh Haghi[a] (a.haghi@alumnos.upm.es), Sandra Martínez-Cuevas[a] (sandra.mcuevas@upm.es), Belén Benito[a]  (mariabelen.benito@upm.es), Juan Carlos Ojeda[b] (jcarlos.ojeda@catastro.minhafp.es).

Affiliations:
a Universidad Politécnica de Madrid. ETSI en Topografía, Geodesia y Cartografía, c/ Mercator 2. E28040 Madrid. Spain.
b Dirección General del Catastro. Pº de la Castellana, 272. E28046 Madrid. Spain.

Corresponding author: Yolanda Torres (y.torres@upm.es)

Abstract

We present a procedure for assessing the urban exposure and seismic vulnerability that integrates LiDAR data with aerial and satellite images. It comprises three phases: first, we segment the satellite image to divide the study area into different urban patterns. Second, we extract building footprints and attributes that represent the type of building of each urban pattern. Finally, we assign the seismic vulnerability to each building using different machine-learning techniques: Decision trees, SVM, logistic regression and Bayesian networks.

We apply the procedure to 826 buildings in the city of Lorca (SE Spain), where we count on a vulnerability database that we use as ground truth for the validation of results. The outcomes show that the machine learning techniques have similar performance, yielding vulnerability classification results with an accuracy of 77% - 80% (F1-Score).

The procedure is scalable and can be replicated in different areas. It is especially interesting as a complement to conventional data gathering approaches for disaster risk applications in areas where field surveys need to be restricted to certain areas, dates or budget.

1. Introduction

Risk is a measure of the expected impact of a natural hazard event on a certain set of exposed, valuable elements. Risk assessment involves the definition of the natural hazard and its intensity in a given territory, the spatial and temporal distribution of assets and other exposed elements and the characterization of the vulnerability of these elements. Results of a risk assessment study are expressed in terms of damage to the population, to the built and natural environments and economic losses. The quality of risk results strongly relies on the availability and quality of data to derive all these inputs (UNISDR, 2015).

Risk assessment is the basis for the development of informed disaster risk reduction activities, which constitute one of the main priorities of the international agenda. In this line, one of the global targets of the UN-promoted Sendai Framework is to increase and make available risk information and assessments to communities and local organizations. Special attention is given to communities lacking resources and infrastructure for collecting data and monitoring risk drivers.

Commonly, collecting data for risk analysis is a resource-consuming task (e. g., Dunbar et al., 2003), becoming unaffordable or unfeasible within research projects or governmental policies for disaster risk reduction. In other cases, the information is existent, yet inaccessible for risk analysts. Thus, there is a need for developing novel procedures that optimize resources to create and/or update natural risk databases in a time- and cost-effective manner.

In this regard, remote sensing techniques are effectively used to obtain relevant, first-order information to analyze the exposed inventory of buildings, to characterize terrain susceptibility and to evaluate the vulnerability of the target area (Pittore et al., 2017). The use of remotely sensed data allows for studying the built environment in detail, given the large amount and variety of mid- and high-resolution data available nowadays. This provides independence from ancillary data sources, such as cadaster databases (Geiss et al, 2015), and enables reducing the time dedicated to field surveys, data validation or expert knowledge acquisition. In fact, the design of field surveys can be optimized with a pre-analysis of the study area with remote sensing techniques.

This study aims at characterizing the built stock of the Spanish city of Lorca by integrating airborne LiDAR points, orthophotos and satellite images to create an exposure and earthquake vulnerability database. We propose a procedure for data integration that intends to be fast and easy to deploy. Similar procedures designed to this end generally follow three major steps: (1) building footprint delineation; (2) attribute extraction from the footprints (such as building floor area and height); and (3) vulnerability allocation to the buildings. The two first steps are addressed with remote sensing while any type of predictive model is used for the third one.

During the last fifteen years, scientists have been using different combinations of remote sensing data to accomplish steps 1 and 2. Building footprints are achieved either by manual digitation (Qi et al., 2017); or through Object-Based Image Analysis (OBIA), i.e., automated image segmentation and classification (Wieland et al., 2012a); or using any available footprint dataset, such as OpenStreetMap (OSM) or cadaster/census databases (Riedel et al., 2015). In some cases, more than one approaches are combined (Wieland et al., 2012b; Geiss et al., 2017); it all depends on data availability or data collection capabilities. For footprint delineation in this paper we have used both, image segmentation and a cadastral database, for comparison. The automated segmentation process usually yields under- or over-segmentation of image objects. In order to improve these results, a more or less sophisticated post-process needs to be applied for segment refinement (Geiss et al., 2017) that increases considerably the time spent in this task. Even after the a posteriori refinement, the complete segmentation process hardly ever provides the desirable polished result. Provided that a perfect footprint is almost impossible to achieve and given that one of our main principles is to keep our procedure as automated as possible, the image segmentation is followed by a rather simple, rapid post-process in a Geographic Information System (GIS).

Next, step 2 is devoted to attribute extraction in each footprint. Attributes such as building floor area and roof material can be extracted from OBIA. However, apart from the segmentation, the subsequent image classification is also an intense process itself, especially in difficult study areas where complex roofs exist with different slopes and materials. Besides, classification requires a sampling process for training and testing datasets compilation that might become overwhelming if large datasets are needed (from dozens to hundreds of samples. E.g. Matsuka et al., 2012). Finally, in some cases a multi-level segmentation and classification process is required (Mück et al., 2013) in order to get a certain level of desired accuracy. Based on these considerations, in this paper we propose to skip the image classification task and extract all the attributes using only the footprints and the LiDAR cloud.

The building height is one of the most difficult attributes to achieve given the limited availability of 3D spatial data. Numerous attempts are done to extract height information using the building shadows present in airborne and satellite images. Traditional approaches require the whole shadow to be completely casted and measurable on the ground (e.g. as applied in Ehrlich et al., 2013) and newer algorithms propose to measure only one segment of the shadow (Su et al., 2015). However, all of them have limitations due to blocked shadows on adjacent buildings or because they are not applicable in slope terrains. To overcome these problems, Sarabandi et al., 2008 developed an algorithms which is not based on shadows. Instead, it only needs the two points of a vertical building edge, but still it cannot be applied in densely populated areas due to building occlusion. Given that, we propose the use of other sources for building height estimation. Typically, 3D data is incorporated from sources such as airborne LiDAR (Ricci et al, 2011), RADAR (Geiss et al, 2015) or Mobile Mapping Systems, MMS (Pittore and Wieland 2013). The latter provides acceptable accuracy, yet MMS data are generally unavailable leading to design specialized data collection campaigns. RADAR data are widely available and its accuracy has been improved over time. To date, we can count on SAR data (Synthetic Aperture Radar) with geometric resolution of 5 m (Geiss et al., 2015) or even 1 m (Polli and Dell'Acqua, 2011). However, LiDAR overcomes MMS in availability and SAR in resolution. LiDAR datasets are becoming increasingly available through national programs such as in Spain (National Plan of Aerial Orthophotography, PNOA) or Italy (Borfecchia et al., 2010), or through tailored projects using LiDAR sensors mounted in unmanned aerial vehicles (UAVs), which are a reality nowadays. The LiDAR data collection is fast and the processing can be totally automated. LiDAR provides intensity information (apart from planimetric and altimetric data) which is useful for material differentiation; and the resolution is higher than SAR as there can be more than one point per square meter.

All these reasons guided us to use LiDAR in this study. The Spanish PNOA provides open access to LiDAR points and orthorectified images. Indeed, the Spanish government through the PNOA former director (personal communication) have expressed their particular interest in exploring all possible applications these data may have. Consequently this research project counts on their support. The use of PNOA data in the field of earthquake engineering that we are presenting here is pioneer in Spain, and could be replicated in other cities.

In other geographic contexts, other authors have worked already with LiDAR for extracting the building height and roof configuration. However either they did not provide any accuracy measure of final building type classification (Costanzo et al., 2016) or they used also images, which provides extra information (geometrical and spectral) about the buildings (Borfecchia et al., 2010, Riedel et al., 2015). In our study, we test a new algorithm called Magic Surface that we specifically developed for LiDAR point classification (Arranz, 2013). Magic Surface outperforms other commercial tools as detailed in section 3.2.1.2.

In this study, the challenge for us is double: (1) to prove that PNOA data is useful for exposure and earthquake vulnerability assessment in a Spanish city; and (2) to prove that the LiDAR point classification with Magic Surface is accurate enough that allows for detecting buildings and extracting vulnerability-related attributes without having to classify the orthophotos.

The study area in Lorca is located in an earthquake-prone region of Spain. Particularly, Lorca was the most affected city after the Mw 5.1 2011 earthquake, involving 9 casualties, over 300 injured people, the temporal reallocation of about 10000 inhabitants, and economic losses of more than 460 M € (Álvarez Cabal et al., 2013). This was the earthquake with the highest impact in Spain in the last 50 years. After the 2011 event, the area of Lorca was the focus of extensive research. Of particular interest are the works by Martínez-Cuevas and Gaspar-Escribano (2016) and Martínez-Cuevas et al. (2017), which provide a detailed building database of different areas of the city that is used as the ground truth data in the present study.

The paper is divided into four main parts. First, we include an initial analysis to determine the statistical relation between the attributes and the earthquake vulnerability in the study area. Second, we present the procedure designed to assess the seismic vulnerability of the built environment. Then, we describe the application of the procedure in Lorca and the results obtained. Finally, we discuss the results, draw some conclusions and introduce future research lines.

2. Previous work: building typologies in Lorca

This section describes the exploratory work that precedes the actual procedure to develop an exposure and vulnerability database from remote sensing data. This is essential for appraising the already-available and pertinent information that can be used to check the quality of the results; and for identifying the target attributes that are required for the vulnerability database.

We study the correlation between the building vulnerability and other building attributes in Lorca, using the building database created by Martínez-Cuevas et al (2017). They analyzed three sample areas of interest in Lorca (Figure 2, right) located in different barrios. The reference database contains 826 buildings whose footprints were taken from the Spanish cadastral database. For each building, the seismic vulnerability is given in terms of the model building types (MBT) as described in the Risk-UE project (Milutinovic and Trendafiloski, 2003).

Six MBT are found in Lorca, one with reinforced concrete structure (RC) and five with masonry structure (M). RC buildings predominate in Barrio San Diego (zone North) and Barrio La Viña, which were built more recently. These buildings have a reinforced concrete frame with unreinforced masonry walls, and thus are associated to the typology RC31 defined in the Risk-UE project. Generally, they are multifamily buildings with more than 3 stories. Very few, dispersed masonry buildings are found in these areas.

On the contrary, the historical city center (zones Center A and Center B) presents a somewhat wider range of building typologies, with a significant presence of masonry MBT (three out of the five masonry types of the city can be found here). In these barrios we find old buildings with unreinforced masonry bearing walls of rubble stone and fieldstone, as well as other unreinforced masonry buildings. The former corresponds to the M1.1 building typology of the Risk-UE project; and the latter correspond to the typology M3.1 of Risk-UE (if the buildings present wooden slabs) or M3.4 (if the slabs are made of reinforced concrete). All these masonry buildings are similar: low rise (usually lower than 3 stories) single-family units without any significant irregularities (plan or vertical). The distribution of these MBT in the study areas is presented in Table 1.

Table 1. (single column) Distribution of MBT in the study areas.

| Nr. Buildings | SPRAWL | | CITY-CENTER | | Total |
|---|---|---|---|---|---|
| | North | Viña | Center A | Center B | |
| M1.1 | 3 | 3 | 16 | 21 | 43 |
| M3.1 | 4 | 0 | 23 | 69 | 96 |
| M3.4 | 2 | 1 | 11 | 25 | 39 |
| RC3.1 | 79 | 360 | 129 | 80 | 648 |
| Total | 88 | 364 | 179 | 195 | 826 |

Martinez-Cuevas et al (2017) grouped the three masonry MBT into one main group due mainly to two reasons: first, the number of buildings of each masonry MBT was not sufficient to create three separate samples in each study area; and second, the three typologies correspond to unreinforced masonry buildings with similar geometric configuration.

According to the literature (e.g. Costanzo et al., 2016) and to our experience in this field, the following attributes commonly have influence in the variable MBT: area, number of stories, and type of roof. In order to confirm this hypothesis, we use the reference database to study the correlation between M and RC building types and these attributes through a contingency table, which informs on the interdependence of a pair of variables (MBT – tested attribute). Five records are removed from the database for this analysis due to missing values.

We conduct a statistical hypothesis testing for each attribute. The null hypothesis $H_0$ is that both variables are independent. For all three attributes, we obtain a p-value = 0.0, meaning that the null hypothesis can be rejected for any level of significance. In other words, the MBT is dependent on the selected attributes. The contingency tables are included as supplementary material to this paper.

In the remaining, we present the procedure followed to calculate these attributes using remote sensing and to infer the MBT of each building.

3. Materials and methods

The procedure designed in this study uses remotely sensed data to obtain exposure and vulnerability databases. It presents three main phases (Figure 1): Phase 1 deals with the stratification of the city in different urban patterns. Phase 2 consists on identifying building footprints and calculating attributes to compose an exposure database. Finally, phase 3 addresses the allocation of vulnerability to each building.

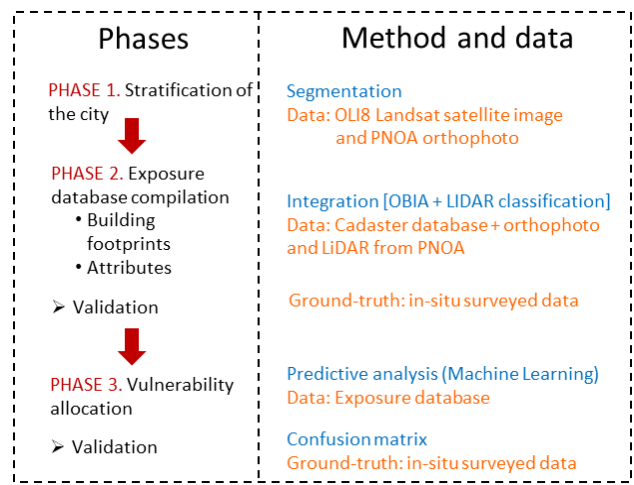

Figure 1. (single column) Phases of the procedure designed to assess the vulnerability of the built environment in Lorca

We use data from different origins ([Dataset] DataLorca). In addition to the ground truth data from the in-situ survey mentioned above, we use remotely sensed data, including satellite images, orthophotos and LiDAR. An 11-band image registered in March 2015 by the Landsat8 satellite (carrying OLI and TIRS sensors for registering 1-9 and 10-11 bands respectively) are used for the city segmentation in phase 1. This image is radiometrically, geometrically, and terrain-corrected, and all bands are 12-bit GeoTiff format. The geometric resolution is 30 m, except for the panchromatic band (band 8) which is 15 m (Landsat8).

We also use in phases 1 and 2 the orthophotos captured in 2013 by the National Plan of Aerial Orthophotography (PNOA) of the Spanish National Geographic Institute (IGN). These 4-band images (RGB + NIR) have a geometric resolution of 50 cm and a planimetric RMSE ≤ 50 cm. In phase 2, the LiDAR point cloud captured in the same photogrammetric flight is included in the analysis. LiDAR point capture is performed with a high-resolution digital camera, equipped with a panchromatic sensor and four multispectral sensors. According to the technical specifications for digital photogrammetric flight with LiDAR, the average density is 1 point/$m^2$. The planimetric and altimetric RMSE for these points are 30 and 20 cm respectively (PNOA).

Finally, we use the building footprints from the Spanish official cadastral database in phase 2.

To integrate all the spatial data in a common geographic environment, we implement a GIS with ArcGIS (ESRI). All spatial information is projected to the reference system ETRS89-UTM30N, which is the Spanish official reference system according to the Royal Decree 1071/2007, of July 27. We work at two scales (Figure 2, left). First, the stratification (phase 1) is conducted in the urban built-up area of Lorca (i.e. excluding the surrounding rural area), with an approximate extension of 9 km$^2$. The number of residential buildings exceeds 6,300 units, according to the cadaster official record, and the number of inhabitants living in these buildings is nearly 60,000. And second, the exposure and vulnerability analysis (phases 2 and 3) are carried out in three smaller sample zones namely, North, Center and South (Figure 2, right) for which we have the ground-truth database. Zone Center covers part of the historical city center, whereas zones North and South correspond to more recent urban developments.

In the next sections, the three phases of the procedure are detailed.

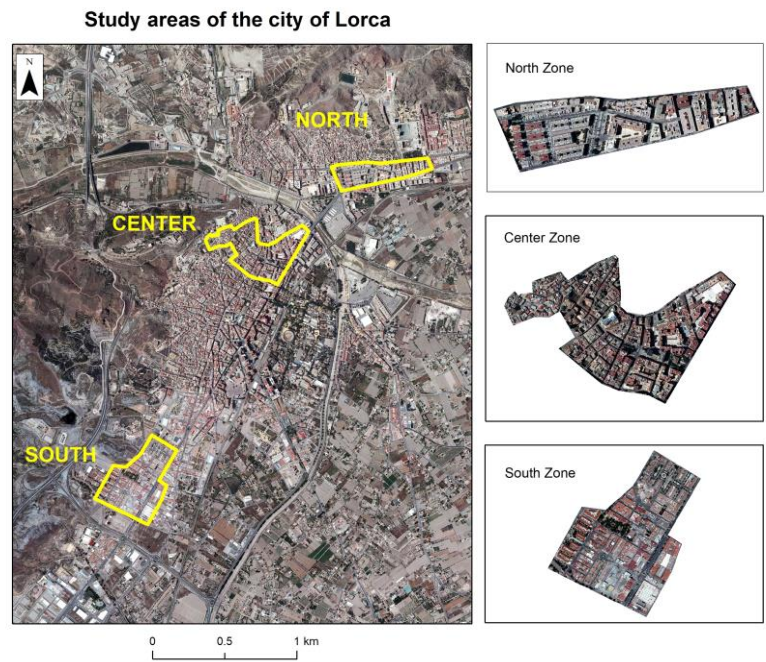

Figure 2. (1.5 column) Study areas considered in the study: the urban center of the city of Lorca for phase 1 (left) and three sample areas for phases 2 and 3 (right).

### 3.1. Stratification of the city

The stratification of the city is the process of dividing the built-up area into regions with homogenous urban patterns (Wieland et al., 2016). These are commonly related to the city growth stages (historic center, middle XX century expansion, recent residential developments) or land use (residential, industrial, recreation). The ensemble of regions with the same urban pattern is a stratum. The stratum is the geographical unit used to perform subsequent stratified machine learning analysis aimed at detecting and characterizing single buildings.

The built-up area layout of Lorca is elongated, with a clear direction SW-NE. Approximately located in the middle, the historical city center presents an irregular urban configuration, as usual in most of the Spanish cities. In the surroundings, we find relatively recent barrios with modern constructions arranged in a more regular street pattern. There is an industrial area located to the SW. Accordingly, we observe six urban patterns in the city, namely: Residential, Historical city center type A, Historical city center type B, Sprawl, Rural, and Industrial.

To stratify the city of Lorca we apply an image segmentation process over the Landsat satellite image. We use the algorithm Watershed as implemented in the software ENVI (Jin, 2012). This algorithm classifies the pixels by increasing greyscale value, starting from the pixels with the lower value. Then, neighboring pixels with similar intensities are grouped to create the homogenous regions, that is, the segments. The Edge method of ENVI is used because the objects that we want to identify have clear boundaries (normally streets delimit the different urban patterns). The result of the segmentation is a polygon layer of segments.

Following the segmentation, the approach of Object-Based Image Analysis (OBIA) includes a classification process. The classification consists of assigning to every segment a label representing its urban pattern using some machine learning algorithm. However, we classify the segments manually at this phase of the procedure given the manageable extent of the city.

## 3.2. Exposure database compilation

The exposure database for this study is a spatial database containing the footprint of each building along with some attributes. The footprints come from two sources: the Spanish cadastral database and the results of an image segmentation process. The first one matches the footprints of the ground-truth. The second ones are detected by a combination of high-resolution orthophotos and LiDAR points from the PNOA project. This second procedure is described in the following sections.

Each building footprint is one entry in the exposure database. Then, new fields are added to the database, which are the building attributes that are related to their earthquake vulnerability.

### 3.2.1. Building footprint delineation

While the cadastral database provides the building footprints directly, extracting the footprints with remote sensing requires a laborious process that is described in the following.

#### *3.2.1.1. Orthophoto segmentation:*

The approach starts with the orthophoto segmentation as described in phase 1 (section 3.1) with the aim of extracting the building roofs. As a result, the polygon layer contains the contour of each object present in the image (roofs, streets, trees, etc.). Within the GIS, the polygons are smoothed and simplified. The simplification is based on the area of the polygon: bigger polygons absorb smaller neighbors. The area threshold value is set after a trial and error process, as explained in section 4.2.1.

As not all the polygons detected correspond to buildings, we carry out a LiDAR point classification procedure to identify the buildings only.

#### *3.2.1.2. LiDAR point classification:*

LiDAR point cloud classification is to assign a class, represented by an integer code, to every LiDAR point. This class refers to the type of object that is reached by the laser (Arranz, 2013). Examples of classes are bare ground, vegetation, and buildings.

The LiDAR classification is done with the algorithm Magic Surface (MS19) implemented in MDTopX software. We developed Magic Surface in 2013 (Arranz, 2013) and it has been improved for this study. The flow chart is presented in Figure 3. The algorithm reads LAZ/LAS files and starts asking for all the parameters. Then, the point classification is done automatically, without the intervention of the user. First, the algorithm works in a grid to search for points with the lowest elevation. These points are classified as bare ground and are compared to their neighbors in a search for differences in Z. Points with differences larger than a threshold set by the user are classified as objects (not bare ground); otherwise, as bare ground. The process is repeated recursively for smaller cell sizes. This preliminary bare ground point classification is refined in three steps: (1) using registration data, (2) searching for peaks in the triangle network (TIN), (3) searching for differences in Z higher than the tolerance, T, defined as a function of the cloud point density (T=1/6 * maximum distance between points). At this point, the final digital terrain model (DTM) is computed.

The algorithm continues with the detection of holes in the DTM. These holes are zones where there are no bare ground points. The projection of the remaining points over these holes allows for identification of higher elements such as vegetation, buildings and other urban furniture. To refine this classification, the algorithm follows two steps: (1) it applies geometric rules related to shape regularity and flatness using the height and area thresholds provided by the user: (2) it makes use of the near infrared band (if available) to further distinguish between buildings and vegetation by means of the Normalized Difference Vegetation Index (NDVI). This final process increases the accuracy of building and vegetation classification significantly. The result is a classified point cloud where bare ground, vegetation (high, medium or low), buildings, bridges (on roads) and urban furniture are distinguished. This classification provides the urban space configuration.

A more detailed explanation of the algorithm Magic Surface is included as supplementary material of this paper.

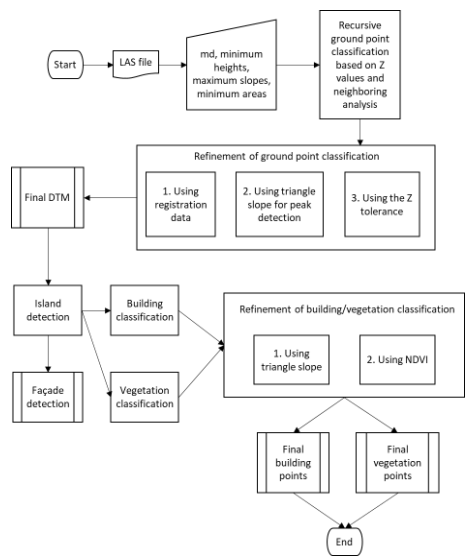

Figure 3. (1.5 columns) Flow chart of the algorithm Magic Surface (MS19).

Magic Surface was tested by Arranz (2013) with very different datasets: urban, rural, and densely vegetated areas, façade of a heritage building, petroglyph, mobile mapping system and a tunnel. All the experiments yielded high accuracies when compared with GNSS or photogrammetric data. Moreover, the algorithm`s performance was tested against others' using an independent reference dataset proposed by the International Society of Photogrammetry and Remote Sensing (ISPRS[1]). The dataset contains 15 sample sites with diverse configuration (urban and rural) and the trial consists of classifying the points and validate the results using the ground truth provided by the ISPRS. The accuracy measures are given in terms of percentage error of the classification of bare ground and object points (ground and object indicator, respectively), and the combination of both (total indicator). The global comparison of algorithms for LiDAR classification promoted by the ISRPS was done in 2003, and since then these datasets are public for testing new algorithms.

The first version of Magic Surface was developed and tested with this independent procedure in 2013, ranking second when compared to other 9 LiDAR classification algorithms (Arranz, 2013). For the present work, we test the improved Magic Surface algorithm again. Results ratify the second place in the global ranking as well as the improvement with respect to the first version. The bare ground, object, and total indicators for all the 15 sample sites are plotted in Figure 4. The object indicator remains below 5% for all samples, except for ID71. The ground classification error is high for 4 samples, exceeding 20% in samples ID11 and ID41. Overall, the total indicator remains below 10% for almost all the samples. The average values for the 15 sample sites of the ground, object and total indicators are 8.0%, 2.6% and 5.4%, respectively.

Figure 5 shows the comparison of average percentage errors obtained by all the algorithms. The current version MS19 presents lower errors than the previous one and gives a total indicator (5.4%) that is very close to the value reported for the top-ranked algorithm (Axelsson, with 4.7%). Of especial importance is the high accuracy in object classification provided by MS19 (only 2.6% of percentage error). In the supplementary material of this paper, we provide other graphs and tables with the percentage errors of all the algorithms in all the sample sites, including Magic Surface (first and last version).

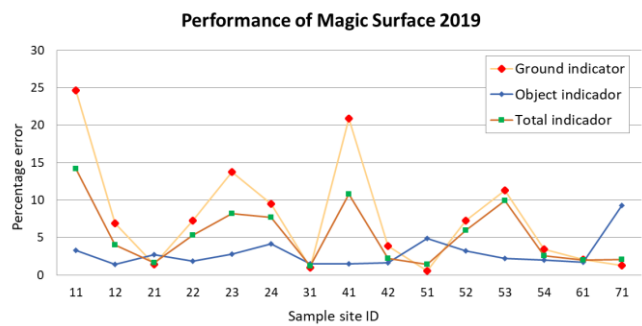

Figure 4. (1,5 columns) Performance of the algorithm Magic Surface (MS19) in terms of percentage error in the detection of ground and object points.

---

[1] https://www.itc.nl/isprs/wgIII-3/filtertest/

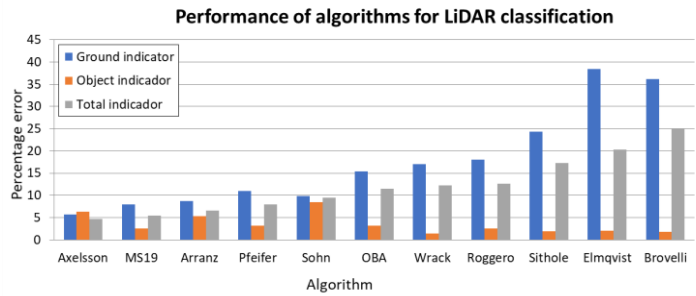

Figure 5. (1,5 columns) Comparison of the percentage errors of all the algorithms for LiDAR point classification that participate in the ISPRS performance trial. MS19 stand for Magic Surface 2019 and Arranz is the first version of magic surface in 2013. The names in the horizontal axis correspond to the other algorithms (Arranz, 2013).

### *3.2.1.3. Integration of the image segmentation and the classified LiDAR*

Within the GIS, the LiDAR points classified as buildings from the previous stage are overlaid on the segmentation layer with the aim of keeping only the segments corresponding to buildings (Figure 6).

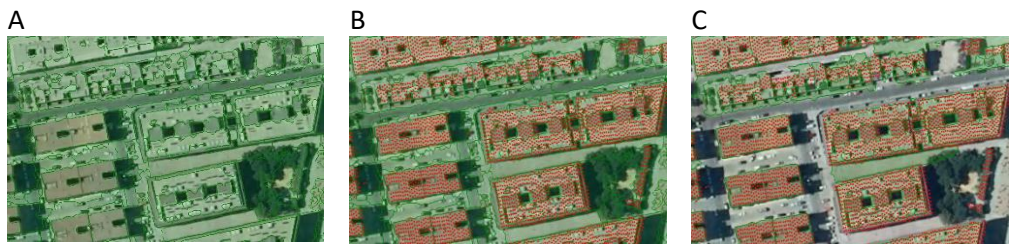

Figure 6. (1.5 columns) Integration of image segmentation and LiDAR. A: Segmentation polygon layer. B: LiDAR points classified as buildings (red dots) superimposed on the segmentation. Points can be seen only on the roofs. C: Selection of segments containing building points.

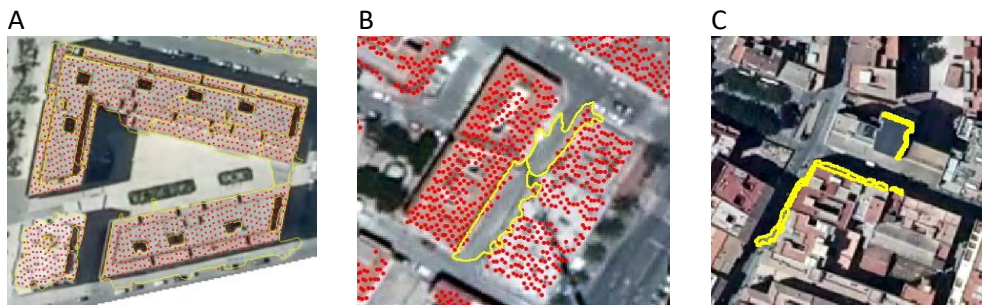

Figure 7. (1.5 columns) A: LiDAR building points creating the actual building footprints, without being affected by heterogeneous elements present on the roof. On the contrary, the polygon layer (yellow contours) shows over-segmentation. B: Some LiDAR building points lie on one polygon belonging to the ground. C: Yellow lines are the contours of very thin, long segments, adjacent to the roofs.

Despite the segmentation parameters are thoroughly tuned, elements located on top of the roofs bias the actual footprint contour (e.g. air conditioning or electrical systems). These elements have a radiometric response that is different from that of the roof surface, thus the segmentation algorithm isolates them into small polygons (Figure 6 A). This is one source of over-segmentation. However, MS19 can be parameterized to disregard these types of noisy elements on the roofs (Figure 7 A). The different coloration of gable roofs due to shadows is another source of over-segmentation, which does not affect the performance of MS19.

The combination of the image segmentation with LiDAR may lead to other issues that distort the delimitation of building footprints:

- Some LiDAR building points are found to be outside the actual building footprints. As a consequence, when superimposing the LiDAR on the segmentation, polygons neighboring buildings, such as streets or trees, will be wrongly selected as buildings (Figure 7 B).
  To avoid this, we calculate the point density for each polygon and select only those polygons with a point density exceeding a certain threshold. Again, the density threshold is established by trial and error (see section 4.2.1 for the actual application).
- Even after applying the point density criterion, some polygons might be wrongly selected as buildings. These are thin, long segments adjacent to buildings where some LiDAR points lie, such as roof eaves or shadows (Figure 7 C).
  To eliminate these polygons, we calculate a shape index $SI = P_a / P_s$, where $P_a$ is the actual perimeter of the polygons and $P_s$ is the perimeter of a square with the same area than the actual polygon. The closer to 1 is SI, the more compact the polygon is. Hence, thin, long polygons will have a low SI value (below a certain threshold) and will be not labelled as buildings. Once more, the SI threshold is obtained after a trial an error process (see section 4.2.1). Besides the SI is calculated, an area-based criterion had to be applied to eliminate small polygons that could not be considered as buildings.

Once all three steps are completed, the result is a geospatial database in which each entry is a refined component of a building footprint. Next, the vulnerability-related attributes are calculated in order to complete the exposure database.

### 3.2.2. Attribute calculation

As exposed in section 2, the attributes required for the exposure database are the area, the number of stories, and the roof type. Also the location of the centroid is included (X and Y coordinates).

The area is calculated for each building footprint using the GIS once the geometry is refined, as well as the centroid location. The number of stories is derived from the building height, which is estimated as the difference of elevation between the roof and the ground. However, the determination of the roof type is more elaborated. We create several prediction models to classify the building roofs into two categories: concrete slab and tile. The machine learning algorithms we use in WEKA are J48, Bayesian networks, and support vector machines (SVM).

J48 is an algorithm to create decision trees adapted for WEKA from the C4.5 algorithm of Quinlan (1993). A decision tree subdivides the training dataset into smaller, purer groups by establishing conditions based on the attribute values. The nodes after a split contain a larger proportion of instances of a certain class (Kuhn and Johnson, 2013).

WEKA allows for setting a key parameter of this algorithm, the confidence factor, f, to regulate the size of the tree. It ranges from 0 for a simpler tree, to 1 for a deeper one.

A Bayesian network is a Directed Acyclic Graph (DAG) represented by means of a set of nodes and arcs (Pearl, J. 1985; Ben-Gal, 2007). The nodes are the variables, i.e., both the building attributes and the MBT. Each arc connects two nodes according to the relationship between them, as parent and child. For each node, a conditional probability of the child is calculated given its parents. Here we use the K2 algorithm implemented in WEKA to create the Bayesian network, tuning the parameter maximum number of parents, p. We also create a TAN (Tree Augmented Network) which is a naïve Bayes network augmented with a tree (Friedman et al., 1997). No parameters have to be set for learning a TAN in WEKA. The a priori probability is calculated using the training data, once discretized. Bayesian networks has proved to be a powerful classifier in this field while being flexible and rather simple (Li et al., 2010; Pittore and Wieland, 2013)

Support vector machines (SVM, Vapnik, 2000) is one of the most flexible and effective machine learning techniques (Kuhn and Johnson, 2013). An SVM calculates the optimal classification surface that separates the samples of different classes, using the attributes of only a subgroup of them (the so-called support vectors). These support vectors are the instances that are closest to the surface, and the separation is called margin. The goal of a SVM is to maximize this margin. The key parameter to be set when using SVM is the cost-parameter, C, that controls the trade-off between the smoothness of the classification surface and the misclassification errors. The smaller the C parameter, the softer the margin; and vice-versa. In the first case, the risk of misclassification is higher but the capacity of the model to generalize is higher too (softer margins tend to avoid over-fitting). We try different kernels with the SVM in WEKA, namely polynomial, Radial Basis Function (RBF), and Pearson VII Universal (PUK, Üstün et al., 2006). The polynomial kernel requires setting the power through the d parameter. The RBF's key parameter is gamma, the kernel width (γ). For the PUK, sigma (s, for the half-width) and omega (o, the tailing factor) have to be adjusted.

In addition to all the attributes calculated up to now (area, centroid location and number of stories) we include the roof slope to create the models. The idea of considering the slope is based on the fact that roofs with low slopes are typically made of a concrete slab, whereas tilted, small roofs are expected to be of tile in our study area. The roof slope is estimated as the median of the slope of the Triangular Irregular Network (TIN) created using the LiDAR points contained within the footprint. The median value is weighted by the triangle area to filter out the influence of small triangles (Eq. 1):

$$(\text{Eq. 1}) \quad S_r = \frac{\sum_{i=1}^{n} S_i \cdot A_i}{\sum_{i=1}^{n} A_i}$$

Where Sr is the roof slope calculated with the n triangles of the TIN; Ai is the area and Si is the slope of each triangle.

Finally, once the attributes are calculated for each building in the database, they are validated using the in-situ survey data. For the area validation, the estimated total built-up area are compared to the total area recorded in the ground-truth database. For the other attributes (number-of-stories and type-of-roof), a confusion matrix is configured to assess the estimation accuracy. Out of all the accuracy measures that can be calculated in a confusion matrix, we present the F1-Score for the sake of simplification, since it involves sensitivity and precision.

### 3.3. Vulnerability allocation

The seismic vulnerability allocation consists on classifying the buildings into one of the two MBT: RC and M. These categories are related to Risk-UE Typologies. Specifically, masonry and RC buildings are related to Risk-UE MBTs M3.1 and RC31, respectively. The classification is carried out through a predictive mathematical model learned from a set of training samples, which relates the vulnerability (dependent variable, DV) with the attributes (independent variables, IV) contained in the exposure database.

We select the machine learning techniques explained for attribute calculation (section 3.2.2) to fit the predictive model. Moreover, we calculate also a logistic regression as applied by Sarabandi et al. (2008) to classify buildings into structural types in southern California. Also Martínez-Cuevas et al. (2017) applied a logistic regression to this ground-truth database in order to relate the MBT to the damage state caused by the Lorca 2011 earthquake. Binary logistic regression predicts the log odds of a building belonging to one MBT [p/(1-p)] as a linear function of its attributes (Kuhn and Johnson, 2013). This classifier has no parameters to tune, what makes its implementation easier. In this study, we use the software SPSS to fit the model.

The predictive models for MBT classification are learned using a training database created by selecting randomly a set of instances from the ground truth database. The remaining instances are left in a testing database that we use for validation of the algorithms' performance.

4. Calculation and results: exposure and vulnerability estimation in Lorca, Spain

Hereunder, we present the application of the procedure to create an exposure and earthquake vulnerability database in the city of Lorca, together with the outcomes of each phase.

### 4.1. Stratification of the city

The stratification of the urban center of Lorca is accomplished by the segmentation of the OLI8 Landsat image (registered in 2015) with parameters scale=30 and merge=80. Afterwards, we classify manually the resulting segments using the PNOA image (captured in 2013) for photo interpretation.

According to the result (Figure 8), the segments classified as *Historical city center A and B* are located in the center of the city, as expected, as well as in the northwest. *Residential* and *sprawl* urban patterns are distributed around the city center, covering a strip from SW to NE; while *rural* and *industrial* patterns are found in the outskirts.

The stratification of the city shows that zones North and South are located in *sprawl* strata, whereas two different strata are identified in the Center: *historical city center type A and type B* (Figure 8). Accordingly, we divide the sample area into two regions, Center A (with medium-sized buildings) and B (with smaller buildings).

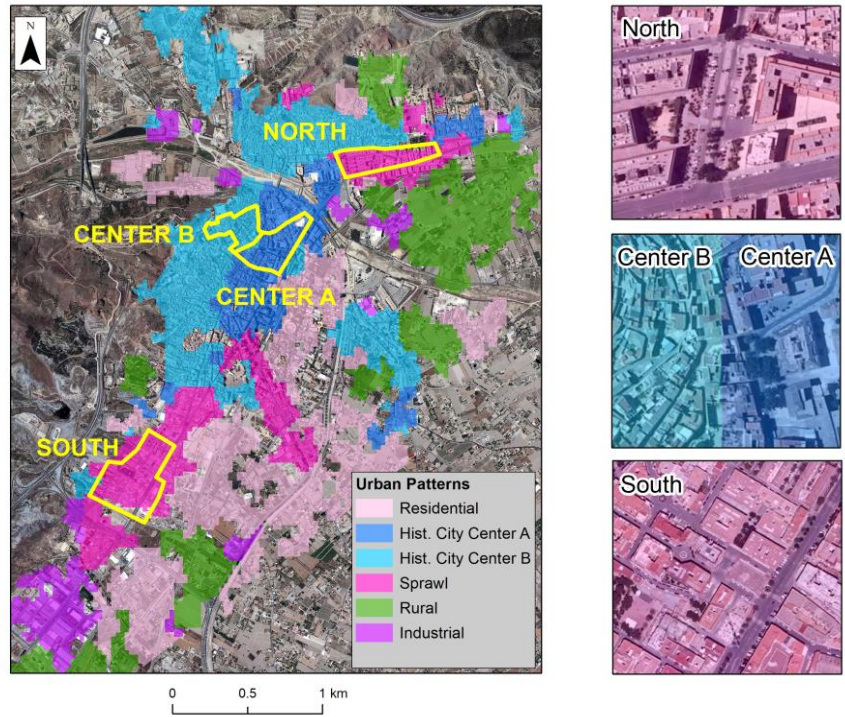

Figure 8. (two columns) Result of the stratification of Lorca into urban patterns.

### 4.2. Exposure database compilation

This section describes the procedure performed to detect buildings and extract their footprints and targeted attributes using a combination of high-resolution orthophotos and LiDAR data. The attributes area, height and centroid location are calculated in each zone. However, for learning the prediction models to estimate the roof type and the MBT, we merge the four zones into two test-beds in order to strengthen the models. Hereunder, we call *Sprawl* the test-bed formed by zones North and Viña (stratum Sprawl); and *City-Center* the test-bed created with the union of zones Center A and Center B.

#### 4.2.1. Building footprint delineation

First, we segment the orthophoto by setting the scale parameter to 60 and the merge parameter to 90 for zones North and Viña. For the two central zones, the parameters are merge=70 and scale=75.

The building contours that result from the segmentation are refined following four fast automated geoprocesses in the GIS: (1) smoothing of segment contours; (2) merging of small segments (with an area lower than or equal to 5 m²) into their larger neighbors; (3) removal of polygons with very few LiDAR points (below a threshold point density of 0.2 points/m$^2$) and (4) removal of thin, elongated polygons (buildings with SI≥1.5 and area≤30 m$^2$). Figure 9 A to D shows some examples of building delineation results.

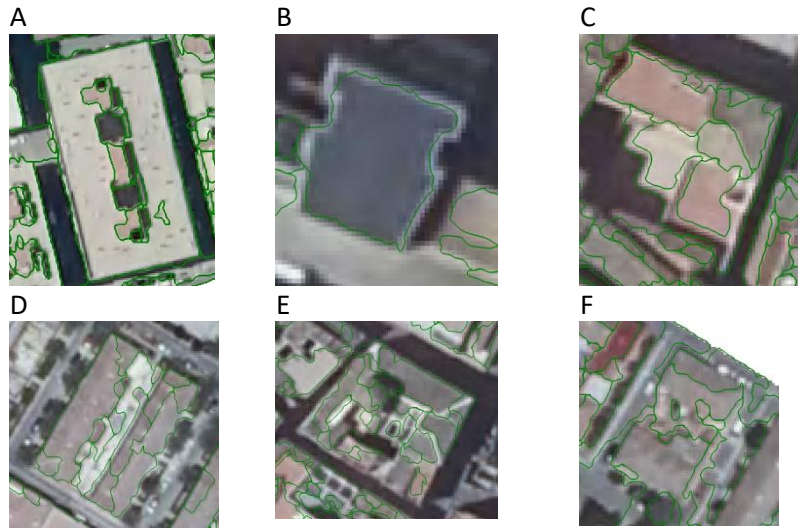

Figure 9. (1.5 columns) Instances of building delineation results in all sample areas. A, B, C, D are instances of acceptable roof segmentations in zones North, Center A, Center B, and South, respectively. E, F are instances of unacceptable roof segmentations in zones Center A and Center B, respectively.

Table 2. (single column) Number of segments obtained as a result of the process here described in each study area compared to the number of buildings in the ground-truth (Ground T.) database.

|  | SPRAWL | | CITY-CENTER | |
| --- | --- | --- | --- | --- |
|  | North | Viña | Center A | Center B |
| Ground T. Nr. Buildings | 88 | 364 | 179 | 195 |
| Estimated Nr. Buildings | 237 | 436 | 1661 | 1037 |

This automated segmentation process tends to over-segment most of the roofs, especially in the two central zones due to their complex morphology (Figure 9 E, F.). Table 2 indicates the number of polygons extracted from the image segmentation, which is 10-fold the actual number of buildings in these central zones. In Sprawl, this difference is still significant, yet less pronounced. Therefore, we cannot take the number of segments directly as the number of building footprints. Some post-processing is required to merge polygons that belong to the same roof (e.g. Geiss et al., 2017). Based on our previous experience, we come to the conclusion that even after long and delicate workflows to this end, a successful result is hard to achieve. Our proposal for such cases is either to dismiss these footprints and get them from another source (official database or digitation, for instance), or to conduct the seismic risk analysis in terms of built-up area instead of number of buildings (Lang, 2013).

Consequently we decide not to use these inaccurate footprints in zones Center A and Center B for MBT estimation. In these zones we work only with the building footprints taken from the cadastral database, whereas in zones North and Viña we work with both footprint datasets to check the impact of this source of uncertainty in the final MBT classification results. Henceforth, the footprints obtained by automated segmentation are referred to as irregular segments/footprints, whilst the footprints taken from the cadastral database are called regular segments/footprints.

### 4.2.2. Attribute calculation

Both, regular and irregular footprints are given the vulnerability-related attributes area, number of stories, roof type and location of the centroid.

Calculating the area of each building and the centroid coordinates is immediate within the GIS. In fact, the area has already been computed for the shape index.

The number of stories is calculated by dividing the building height by the average inter-story height, which we set in 3 m in the study area based on our expert knowledge. In this work, the building height is obtained by subtracting the elevation of the roof LiDAR points and the ground elevation given by the DTM. The DTM is previously derived for each study area in MDTopX using the LiDAR points classified as ground. To this end, the LiDAR point cloud is previously classified with Magic Surface. The combination of key parameters to tune the algorithm that yields the most satisfactory classification is shown in Table 3. The maximum distance between points is set in 3m.

Table 3. (single column)Optimal combination of key parameters to classify the LiDAR points of the four zones with Magic Surface (MS19). mHHV, mHMV, mHLV stand for minimum height for high-, medium- and low-vegetation, respectively; MHV stands for maximum height of vegetation; mAB stands for minimum area of buildings. All values are given in meters.

| Zone | mHHV | mHMV | mHLV | MHV | mAB |
|---|---|---|---|---|---|
| North | 4 | 1.5 | 0.5 | 12 | 50 |
| Center A | 3 | 1.5 | 0.5 | 10 | 30 |
| Center B | 3 | 1.5 | 0.5 | 10 | 30 |
| Viña | 5 | 1.5 | 0.5 | 20 | 50 |

We add to the exposure database the height category, an attribute that depends on the number of stories and it is used in the definition of the MBT. Three categories are considered according to Risk-UE: low-rise (L-rise), for buildings with 1 or 2 stories; mid-rise (M-rise), for buildings with 3 to 5 stories; and high-rise (H-rise), for buildings with more than 5 stories.

We calculate the prediction models for estimating the type of roof as explained in section 3.2.2. Table 4 summarizes the number of instances of each MBT randomly selected for training and testing in each test-bed. For the irregular footprint datasets, we only consider segments with area larger than 50 m$^2$ according to the typical minimum area of the buildings in this zone. Smaller polygons would bias the analysis. A total of 15 and 12 regular instances with missing values are discarded from the regular training datasets for Sprawl and City-Center test-beds, respectively.

Table 4. (single column) Size of the training and testing datasets for roof type classification in the two test-beds. R.F and I.F stand for regular and irregular footprints, respectively. C. Slab stands for concrete slab roof.

|  | Nr. Samples | | | | |
|---|---|---|---|---|---|
|  | Training dataset | | Testing dataset | | |
|  | C. Slab | Tile | C. Slab | Tile | Total |
| SPRAWL – R.F | 50 | 50 | 147 | 190 | 437 |
| SPRAWL – I.F. | 30 | 30 | 213 | 35 | 308 |
| CITY CENTER – R.F. | 50 | 50 | 48 | 214 | 362 |

With the ReliefF algorithm implemented in WEKA (Kononenco, 1994), we rank the attributes to get an insight into their discriminant power. The height, centroid location (x and Y coordinates), and roof slope are the top-four attributes in the three test-beds. We select the optimal parameters of all algorithms by means of a K-fold cross-validation approach, with k=10 for regular footprints and k=6 for irregular ones. The parameters range as follows: f = {0.05…1}, p = {1,2}, C={1…20}, d={1,2,3}, γ={0.01…50}, s={0.5…2}, o={0.5…10}. Table 5 shows the results of the three most (a priori) accurate algorithms ordered by decreasing average F1-Score and kappa statistic, κ, along with their optimal parameters. In principle, all the learning machines created are able to predict the roof type with very high accuracy, with F1-Scores over 80% in most cases.

Table 5. (1.5 column) Ranking of the most accurate learning machines obtained in the training stage for roof type classification. NormPolyK stands for normalized polynomial kernel.

|  | SPRAWL - regular footp. | | |
|---|---|---|---|
| Learning Machine | F1-Score | kappa | Parameters |
| SVM (NormPolyK) | 85% | 0.7 | C=10, d=2 |
| Bayes Net K2 | 81% | 0.62 | p=2 |
| J48 Decision Tree | 79% | 0.58 | f=0.25 |
|  | SPRAWL - irregular footp. | | |
| Learning Machine | F1-Score | kappa | Parameters |
| SVM (PUK) | 95% | 0.9 | C=10, o=5, s=1 |
| J48 Decision Tree | 90% | 0.8 | f=0.50 |
| Bayes Net K2 | 88% | 0.77 | p=1 |
|  | CITY CENTER - regular footp. | | |
| Learning Machine | F1-Score | kappa | Parameters |
| SVM (PUK) | 84% | 0.67 | C=10, o=1, s=0.5 |
| J48 Decision Tree | 80% | 0.6 | f=0.50 |
| Bayes Net K2 | 70% | 0.4 | p=2 |

In agreement with the attribute ranking given by ReliefF, the decision tree uses the attributes height and centroid location for the Sprawl area; and height and slope for the City-Center. The attribute roof slope does not seem to play an important role when inferring the roof type in this study. However, we recommend to include it in the exposure database since it involves a rather simple calculation (once the LiDAR cloud is classified) and it might be useful for other study areas. Moreover, the roof slope is a very important attribute in other natural-risk related situations, such as roof-top snow/ice/volcanic ash accumulation.

*Validation*

The area is validated only for irregular segments, since the regular segments are taken from the same source as the ground truth and the agreement will be total. To carry out this verification, the total footprint area estimated in each study area is compared with the total area of the ground truth footprints. For this comparison, only the segments with an area larger than 20 m² have been included, since smaller polygons cannot be considered as buildings. The values of observed and estimated floor areas are given in Table 6 along with the percentage error (P.E.) both for each zone and for the joint test-beds. In zones North and Center B, the P.E. is lower than 1%, indicating an excellent performance of the procedure here designed for built-up mask estimation using image segmentation followed by a rather simple GIS analysis in these areas. In zone Viña we find a P.E. of 1.3%, also very satisfactory. Zone Center A presents the highest P.E., reaching 7%, due to the complexity of these roofs. This constitutes a challenge for image segmentation.

Overall, the P.E. for Sprawl test-bed is 1% and for City-Center is 3.6%. This means that the method presented here would estimate an area of 104 m2 for a building with an actual area of 100 m2. This difference is negligible when classifying buildings into construction typologies. The result is an accurate built-up mask of the study areas.

Table 6. (single column) Validation of the attribute area. P.E. is the percentage error. Area values are given in m$^2$.

| Test-bed | SPRAWL | | CITY CENTER | |
|---|---|---|---|---|
| Zone | North | Viña | Center A | Center B |
| Area of cadastral Footprints | 39458.55 | 72813.26 | 51912.75 | 19844.49 |
| Area of extracted Footprints | 39737.59 | 73802.65 | 48470.19 | 19830.49 |
| P.E. in each zone | 0.70% | 1.34% | 7.10% | 0.07% |
| P.E. in test-bed | 1.02% | | 3.59% | |

To validate the number of stories, we compare our estimation in each footprint with the corresponding building in the ground-truth database. Figure 10 shows the histograms of the residuals of the number of stories. In all cases, we successfully obtain a majority of residual zero. In Sprawl, a correct prediction is done for 378 buildings (86%) for regular footprints. The remaining 60 buildings have an error of only 1 story (≤3 m). When using the irregular footprints, the accuracy slightly decreases. The height is correctly predicted for 215 footprints (69%) while 65 (21%) have an error of 1 story. In City-Center, 274 height estimations are correct: an error of 1 story is made in 66 buildings (18%). The maximum error in building height estimation reaches 4 stories in the irregular footprints of Sprawl and 3 stories for City-Center. Nevertheless, in both cases these errors only represent the 10% of the footprints. These satisfactory results prove the effectiveness of MS19 for LiDAR classification.

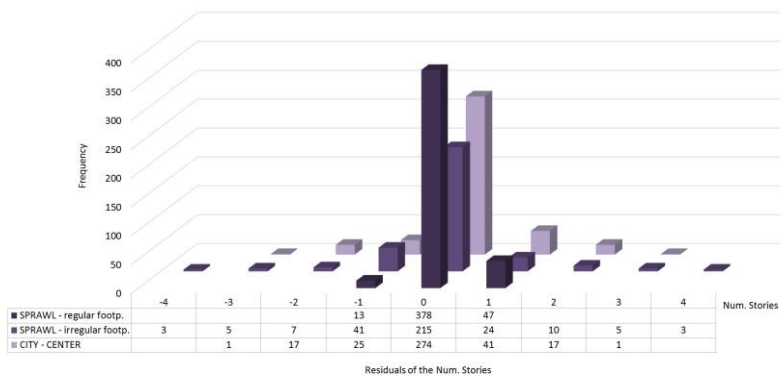

Figure 10. (1.5 column) Histograms of the residuals of the num. stories in each test-bed.

Finally, the type of roof is also verified using the remaining ground-truth samples that were used to create the testing datasets (Table 4). Table 7 lists the average F-1 Score for the three test-beds and Table 8 presents the confusion matrices obtained with the predictive model with higher F1-Score.

If compared with the a priori accuracy estimation (Table 5), SVM with PUK kernel generalizes well as it provides the highest accuracies in Sprawl for irregular footprints and in City-Center. However, SVM with a normalized polynomial kernel is outperformed by the Bayesian network with K2 algorithm in Sprawl for regular footprints, which yields an average F1-Score close to 90%. Also J48 outperforms SVM in that test-bed. Actually, the accuracies of the Bayesian network and of the decision tree improve a 7% with respect to the a priori evaluation. Overall, the average accuracy is high, especially when working with regular footprints. It is worth to notice that the models fit for irregular footprints, which present the highest a priori accuracy, perform worse when evaluated in a new dataset.

Table 7. (single column) Average F1-Score of the roof type classification obtained with the three selected learning machines. R.F. and I.F stand for regular and irregular footprints, respectively

| SPRAWL (R.F.) – F1-Score | |
|---|---|
| Bayes Net K2 2P | 88% |
| J48 Decis. Tree | 86% |
| SVM (NormPolyK) | 83% |
| SPRAWL (I.F.) – F1-Score | |
| SVM (PUK) | 76% |
| Bayes Net K2 1P | 72% |
| J48 Decis. Tree | 69% |
| C.CENTER (R.F.) – F1-Score | |
| SVM (PUK) | 81% |
| J48 Decis. Tree | 77% |
| Bayes Net K2 2P | 74% |

Table 8. (single column) Confusion matrices of the testing phase for roof type classification. R.F. and I.F stand for regular and irregular footprints, respectively. C. Slab stands for concrete slab.

| SPRAWL - R.F. | | Predicted: BAYES K2 2P | |
|---|---|---|---|
| | | C.Slab | Tile |
| Actual | C.Slab | 121 | 26 |
| | Tile | 15 | 175 |
| SPRAWL - I.F. | | Predicted: SVM PUK | |
| | | C.Slab | Tile |
| Actual | C.Slab | 44 | 4 |
| | Tile | 32 | 182 |
| C.CENTER - R.F. | | Predicted: SVM PUK | |
| | | C.Slab | Tile |
| Actual | C.Slab | 158 | 55 |
| | Tile | 2 | 33 |

Once the exposure database is complete and its attributes are verified, we use them to calculate the machine learning models for allocating the vulnerability to the buildings present in the test-beds.

### 4.3. Vulnerability allocation

Vulnerability allocation consists on assigning an MBT to each building in the exposure database. In the following, we describe the process followed to learn the predictive models, to classify the footprints, and to validate the results.

#### 4.3.1. Predictive model calculation and building classification

Table 9 shows the number of training and testing samples that are available to create the predictive models for regular and irregular footprints. 8 out of the 12 instances lost due to missing values when predicting the roof type have been recovered for this MBT classification. The scarcity of M buildings in the Sprawl test-bed as described in section 2 does not allow for testing the predictive model in an unseen dataset. In fact, we decide not to work with the irregular footprints due to the small number of instances of M structures (6) that limits the training dataset size. Hence, for the regular footprints of Sprawl we calculate the models and rank them according to the a priori accuracy estimated through a k-fold cross-validation. If a testing dataset is available in the future, it will be possible to check the generalization capability of the proposed models.

Table 9. (single column) Number of samples in the training and testing datasets for MBT classification. R.F and I.F. stand for regular and irregular footprints, respectively.

|  | Nr. Samples | | | | |
|---|---|---|---|---|---|
|  | Training dataset | | Testing dataset | | |
|  | RC | M | RC | M | Total |
| SPRAWL – R.F | 14 | 14 | 409 | 0 | 437 |
| SPRAWL – I.F. | 6 | 6 | 296 | 0 | 308 |
| CITY CENTER – R.F. | 60 | 60 | 93 | 157 | 370 |

With the training samples randomly selected, we learn different predictive models for the regular footprints using the learning machines explained in section 3.3. With the ReliefF algorithm we rank the attributes to have a first idea of their discriminant power. The building height and centroid (X,Y) coordinates are the first three attributes of the raking for Sprawl and City-Center, followed by the roof type.

In the training stage, a k-fold cross-validation strategy is implemented aimed at searching for the optimal parameters, with k=1 for City-Center and k=4 for Sprawl. To this end, we try different combinations of values for all the parameters that the algorithms require, as described in section 4.2.2 for attribute calculation. For the logistic regression, we apply a forward stepwise method for the selection of variables that are considered to create the model. Table 10 ranks the most accurate predictive models created according to their average F1-Score and kappa statistic, κ.

Table 10. (1.5 column) Ranking of the most accurate learning machines obtained in the training stage for MBT classification.

|  | SPRAWL - regular footp. | | |
|---|---|---|---|
| Learning Machine | F1-Score | kappa | Parameters |
| SVM (PUK) | 89% | 0.79 | C=5, o=2, s=0.2 |
| J48 Decision Tree | 89% | 0.79 | f=0.25 |
| Logistic Reg. | 89% | | |
| Bayes Net TAN | 79% | 0.57 | |
|  | CITY CENTER - regular footp. | | |
| Learning Machine | F1-Score | kappa | Parameters |
| Logistic Reg. | 84% | | |
| J48 Decision Tree | 84% | 0.68 | f=0.05 |
| Bayes Net TAN | 82% | 0.63 | |
| SVM (PUK) | 81% | 0.62 | C=1, o=1, s=1 |

The decision trees obtained are very simple, as they consider only the attributes height and the X coordinate of the centroid. The logistic regression creates a model in which the same two attributes are initially included, followed by the roof type. This is in agreement with the ReliefF rank of attributes.

All the models present a high a priori accuracy, with F1-Scores over 80%. We predict the MBT of the instances in the testing dataset of City-Center (Table 9) to check the generalization capacity of the models fit for this test-bed.

### 4.3.2. Validation of classification results:

To validate the results of the MBT classification in zone City-Center, we calculate the confusion matrices for the four selected predictive models and the average F1-Score. Table 11 shows the ranking of prediction models. This ranking of performance matches the ranking of a priori accuracy (Table 10), with a small reduction of 5% in F1-Score. This is an indicator of the robustness of the learning machines implemented, which are able to generalize without any significant loss of accuracy.

Table 11. (single column) Average F1-Score of the MBT classification obtained with the four selected learning machines.

| Predictive model | F1-Score |
|---|---|
| Logistic Reg. | 80% |
| J48 Decis. Tree | 79% |
| Bayes Net (TAN) | 77% |
| SVM (PUK) | 77% |

Table 12. (single column) Confusion matrices of the MBT classification with the four selected learning machines

|  |  | Predicted: Logistic Reg | |
|---|---|---|---|
|  |  | RC | M |
| Actual | RC | 62 | 31 |
|  | M | 14 | 143 |
|  |  | Predicted: J48 Dec. Tree | |
|  |  | RC | M |
| Actual | RC | 67 | 26 |
|  | M | 24 | 133 |
|  |  | Predicted: Bayes Net (TAN) | |
|  |  | RC | M |
| Actual | RC | 67 | 26 |
|  | M | 29 | 128 |
|  |  | Predicted: SVM (PUK) | |
|  |  | RC | M |
| Actual | RC | 57 | 36 |
|  | M | 15 | 142 |

## 5. Discussion and conclusions

In this paper we present a comprehensive procedure to integrate LiDAR and satellite/aerial imagery to create an exposure and earthquake vulnerability database. The ultimate goal is to reduce the time and cost of collecting this type of data by using remote sensing and machine learning techniques, together with expert knowledge. This procedure can be applied to an entire study area or in combination with other traditional techniques in sub-areas where field-surveys are restricted or not feasible (due to issues such as rugged terrain, dangerous zones, dense urban areas, etc.).

We work in two different urban areas of the city of Lorca (Spain). One is located in the historical city center and the other one in a sprawl area with a more recent development. Through SVM, decision trees, binary logistic regression and Bayesian networks we accurately classify the buildings into two building structural types, namely masonry and reinforced concrete. The former is linked to the Risk-UE MBT M3.1 while the latter is RC31.

The results obtained with this procedure allow for creating an initial seismic vulnerability database. Each entry of the database stores the geometry of the footprint along with the following set of attributes: (X,Y) coordinates of the centroid, floor area, number of stories, height category, roof slope, roof material and shape index. This is already useful for earthquake risk evaluation. Of course this initial classification could be complemented in a further phase to subdivide the masonry MBT and/or to include new attributes derived from those, such as direction, other shape indices or the location within the block.

Regarding the procedure proposed, we can highlight that:

- The differentiation of phase 2 and 3 allows for keeping the exposure database separated from the seismic vulnerability allocation. This partial result is interesting and useful by itself, since it constitutes the basis of the exposure component for any disaster risk analysis, regardless of the type of hazard under consideration.

- The integration of LiDAR in the procedure here developed enables the selection of segments belonging to building roofs after the segmentation as an effective way to generate the built-up mask. This saves the cumbersome image classification process.
- The initial stratification of the city into strata where the buildings are similar and share contextual features, allows for keeping the feature vector dimensionality low. This is essential for learning simple and powerful predictive models, such as logistic regression, decision trees or Bayesian networks. Working in homogeneous strata also allows for the fine-tuning of the parameters considered in LiDAR classification, image segmentation, and predictive modeling.
- To get a suitable LiDAR classification is a tedious and delicate task that needs to be done before using the datasets. Magic Surface (MS19) simplifies this process as it provides a rapid, truthful automated point classification, which is reflected in the accurate outcomes of the built-up mask and building height estimation. A good classification is essential, since the rest of the steps depends on it.
- The image segmentation is a challenging process in this study, particularly in the center areas where the buildings present complex gable roofs. In most of the cases the final footprints are composed by several polygons. This multipart entities are conceptually difficult to handle when creating the predictive models.
- All machine learning algorithms applied in this study yield very similar final classification accuracies, with differences in F1-Score ranging from 3% to 7% in the different tests. However, there is a significant difference between them with respect to parameter setting. SVM is a more flexible classifier, yet it requires more parameters to be set. On the contrary, the rest of the classifiers used here (logistic regression, Bayesian networks and decision trees) are able to provide similar o even better accuracies than a SVM while involving a simpler and shorter setting process.
- The availability of open access data is crucial to accomplish the project. The promotion of open distribution policies is key for numerous research fields in which spatial data analysis is involved.
- One of our main goals is to keep the procedure fast and easy to deploy. To this end, we decided not to classify the orthophotos and not to conduct long post-processing tasks. The whole process described here for analyzing 826 buildings (including data download and pre-process, LiDAR classification, image segmentation, data integration within the GIS, and predictive modelling) has been applied by two skilled individuals in five working days.

Regarding the final results and their uncertainty, the quality of the MBT classification results depends on several factors:

- The segmentation for footprint extraction. The segmentation seems to be the factor that introduces the highest uncertainty. Heterogeneous areas, with a great variety in footprint size and shape, are poorly assessed by the segmentation algorithm. This is apparent in the City-Center test-bed, where the segmentation yields a number of polygons significantly different from the current number of buildings. Consequently, the resulting segmentation footprints cannot be used in the rest of the study. In Sprawl, a comparison of regular and irregular footprints is done for roof type estimation. With the irregular footprints, the final F1-Score obtained in the testing stage decreases around a 20% with respect to the a priori values (Table 4 and Table 6). When using the regular shapes from the cadaster, the testing accuracy is very similar to the accuracy of the training phase, or even higher. This seems to indicate that the heterogeneity of the attribute values affects the predictive models performance for irregular footprints.
- The LiDAR classification propagates uncertainty as it is used for footprint selection and building height estimation. Nevertheless, this uncertainty is not very relevant as the results of such processes are very accurate (see Table 6 for validation of the area the polygons selected to create the footprints and Figure 10 for building height validation).
- The predictive models configured. We have used four classifiers with different mathematical approach in order to avoid a possible bias due to the use of one single algorithm. In general terms, there are no remarkable differences in the behavior of all the learning machines configured in the experiments conducted.

Finally, we find some items that should be addressed in future research:

- With the post-segmentation processes that are available to date it is not feasible to get one-polygon footprints in case of complex roofs. To avoid manual digitization or if official building databases are not reachable, then further research needs to be done either to design a segmentation algorithm able to create one single polygon out of one roof (even if the roof is gabled o has elements over it, such as chimneys or air conditioning equipment) or to merge all the segments belonging to the same roof.
- As the over-segmentation does not allow for counting the actual number of roofs, we propose to work with the built-up area (in $m^2$) of each MBT instead of with the number of buildings. The whole seismic risk study can be done in terms of built-up area, which enables to overcome the problem of over-estimating the number of buildings.
- The procedure presented here for exposure and vulnerability assessment is rapid and accurate. The predictive models created could be used to extend the study to the whole city of Lorca. Moreover, this procedure could be applied in other Spanish cities using the PNOA data or in other countries where similar datasets are available.

Conflicts of interest: none

Acknowledgements: Authors want to thank the PNOA programme of the Spanish National Geographic Institute for making available the spatial data used in this study.

Funding: This work is part of the MERISUR Project: Methodology for an Effective RISk assessment of URban areas, funded by the Spanish Ministry of Economy and Competitiveness, National Program for Research, Development and Innovation oriented to Societal Challenges, reference CGL2013-40492-R.